\documentstyle[aaspp4m,epsf]{article}
\newcommand{\figcomment}[1]{#1}
\def\unsetyr{\def\oyear{\relax}\def\cyear{\relax}\def\cyeara{a\relax}\def\cyearb{b\relax}\def\cyearc{c\relax}\def\cyeard{d\relax}}
\def\setyr{\def\oyear{(}\def\cyear{)}\def\cyeara{a)}\def\cyearb{b)}\def\cyearc{c)}\def\cyeard{d)}}
\unsetyr
\def\jcite#1{\setyr\cite{#1}\unsetyr}

\def\rmmat#1{{\hbox{\rm #1}}}
\def\rmscr#1{\rmmat{\scriptsize #1}}
\newcommand{\be}{\begin{equation}}
\newcommand{\ee}{\end{equation}}
\newcommand{\bt}{\begin{table} \begin{center}}
\newcommand{\et}{\end{center} \end{table}}
\newcommand{\ba}{\begin{eqnarray}}
\newcommand{\ea}{\end{eqnarray}}
\newcommand{\ie}{{\it i.e.~}}
\newcommand{\cf}{{\it c.f.~}}
\newcommand{\eg}{{\it e.g.~}}
\newcommand{\comment}[1]{\relax}
\def\eqref#1{Equation~\ref{eq:#1}}

\def\figref#1{Figure~\ref{fig:#1}}

\def\tabref#1{Table~\ref{tab:#1}}

\def\ired{I\!\!\!\rmmat{-}}

\begin{document}
\title{Gravitational Radiation from Strongly Magnetized White Dwarfs}
\author{Jeremy S. Heyl}
\authoremail{jsheyl@tapir.caltech.edu}
\affil{
Theoretical Astrophysics,
mail code 130-33,
California Institute of Technology,
Pasadena, CA 91125}
\begin{abstract}
The magnetic fields of white dwarfs distort their shape generating an
anisotropic moment of inertia.  A magnetized white dwarf which rotates
obliquely relative to the symmetry axis has a mass quadrupole moment
which varies in time, so it will emit gravitational radiation.  LISA
may be able to detect the gravitational waves from two nearby, quickly
rotating white dwarfs.
\end{abstract}

\section{Introduction}

Compact binaries consisting of a white dwarf and a late main-sequence
star such as AM~Her and DQ~Her cataclysmic variables (CVs) are one of
the primary sources of gravitational radiation observable by the LISA
mission (\cite{LISADOC}).  As the two stars orbit each other with a
period of a few hours, the energy of their orbit is slowly lost to
outgoing gravitational waves.  Additionally, the white dwarfs
themselves are rotating.  For AM~Her binaries, the white dwarf
rotation period is approximately synchronized with the orbital period.
The white dwarf in a DQ~Her binary rotates ten or more times faster
than the binary orbit.  In particular, the DQ-Her, white dwarfs AE
Aquarii and WZ Sagittae rotate with a periods of 33 and 28 seconds
respectively.  If these two stars have a significant quadrupole moment
that is not aligned with the rotation axis, they will be sources of
observable gravitational radiation.

Several authors have examined whether quickly rotating white dwarfs 
on their own will be important sources of gravitational waves.
\jcite{1999gr.qc.....3042L} explored the possibility that
gravitational waves can drive an instability in the r-modes
of near-critical rotating white dwarfs analogous to process that
\jcite{1983PhRvL..51...11F}, \jcite{1987ApJ...314..234C}, 
\jcite{1998PRD..58.084020O}, \jcite{1998PRL..80.4843L}, 
\jcite{1999ApJ...510..846A} have argued may operate in rapidly
rotating neutron stars.  Lindblom found that this instability would
take several gigayears to grow in the observed DQ Her objects.
Furthermore, the r-mode instability can only be excited in hot
($T>10^6$~K), massive white dwarfs ($M>0.9~\rmmat{M}_\odot$) rotating within
twenty percent of their critical period; therefore, it is unlikely
that this instability plays an important role in white dwarfs observable
by LISA.

The magnetic field of neutron stars may induce a sufficiently large
quadrupole moment to affect their spin-down (\cite{Shap83}) and to
produce observable gravitational radiation
(\cite{1996A&A...312..675B}, \cite{Brad98}).  The internal magnetic
fields of white dwarfs distort their structure similarly (\eg
\cite{1968ApJ...153..797O}).  If a quickly rotating white dwarf
harbors a strong internal magnetic field, it may be an important
source of gravitational radiation.  

The fastest rotating white dwarfs have periods of several tens of
seconds while typical white dwarfs rotate in about hour.  This places
them in the most sensitive region of the spectrum for the LISA
interferometer.

\section{Magnetically Induced Distortion}

The magnetic field supplies an anisotropic pressure.  The
tensor virial theorem quantifies the effect of the field on the structure
of the star.  If the field points along 
the $z$-axis and the total magnetic energy (${\cal M}$) is much less
than the internal energy ($\Pi$),
\be
\frac{W_{zz}}{W_{yy}} = 1 - 2 \frac{{ \cal M}}{\Pi} = 
	1 - 6 \frac{{\cal M}}{|W|}.
\label{eq:virrat}
\ee
where $W_{ij}$ is the potential energy tensor (\cf \cite{Binn87}), 
$W=W_{ii}$ is the total potential energy of the star, and $z$ is the
symmetry axis of the magnetic configuration 

The lowest order term in a multipole expansion of the gravitational
radiation in the wave zone (\ie $d \gtrsim c/\Omega$) of a slowly
moving source is the quadrupole term.  Because
general relativity only modifies the structure of white dwarfs
slightly, the reduced quadrupole moment
($\ired_{ij}$) is well approximated by the Newtonian limit (\cite{Misn73})
\be
\ired_{zz} = \int \rho \left ( z^2 - \frac{1}{3} r^2 \right ) d^3 x + {\cal O}\left (\frac{G M}{c^2 R} \right ).
\label{eq:ireddef}
\ee
If one assumes that the internal field is uniform and that the isodensity 
surfaces of the distorted star are similar ellipsoids (\cite{Binn87}),
this yields 
\be
\ired_{zz} \approx -5 \frac{{\cal M}}{|W|} I_0
\label{eq:vir}
\ee
where $I_0$ is the moment of inertia of the unperturbed spherical
star.  A negative value of $\ired_{zz}$ indicates that the
configuration is oblate.

\jcite{1953ApJ...118..116C} treated a similar problem by minimizing
the total energy of a uniform density sphere with a uniform 
internal field and a dipolar external field by perturbing its surface 
axisymmetrically.  They obtained
\be
\ired_{zz} \approx -\frac{9}{2} \frac{{\cal M}}{|W|} I_0
\label{eq:pert}
\ee
for small quadrupolar deformations.

Detailed modeling indicates that the distortion also depends on the
structure of the internal magnetic field.  Specifically stars with a
larger fraction of toroidal fields tend to be prolate while stars
whose fields are dominated by a poloidal component are oblate.  The
distorted structure of the star is calculated by assuming that the
equation of state is strictly barotropic, that both the magnetic field
and rotation only slightly perturb the structure of the star and that
the current and the field vanish at the surface
(\cite{1960ApJ...131..227W}, \cite{1961ApJ...133..170W},
\cite{1968ApJ...153..797O}).  The conditions specify the field through
an eigenvalue equation.  

The \jcite{Chan39} equation of state (with $\mu_e=2$) determines the
unperturbed structure of the star.  This equation of state provides an
adequate approximation for this purpose and is relatively simple
analytically compared to equations of state which include Coulomb and
other corrections (\eg \cite{1961ApJ...134..683H}).

\figref{models} shows how the reduced quadrupole moment depends on the
structure of the field and the central density of the white dwarf.  As
the toroidal field increases relative to the poloidal field, the star
becomes more prolate.  Additionally, as the central density
increases the white dwarf becomes more centrally concentrated, and the
magnetic field is less effective at distorting the star.
\figcomment{
\begin{figure}
\plottwo{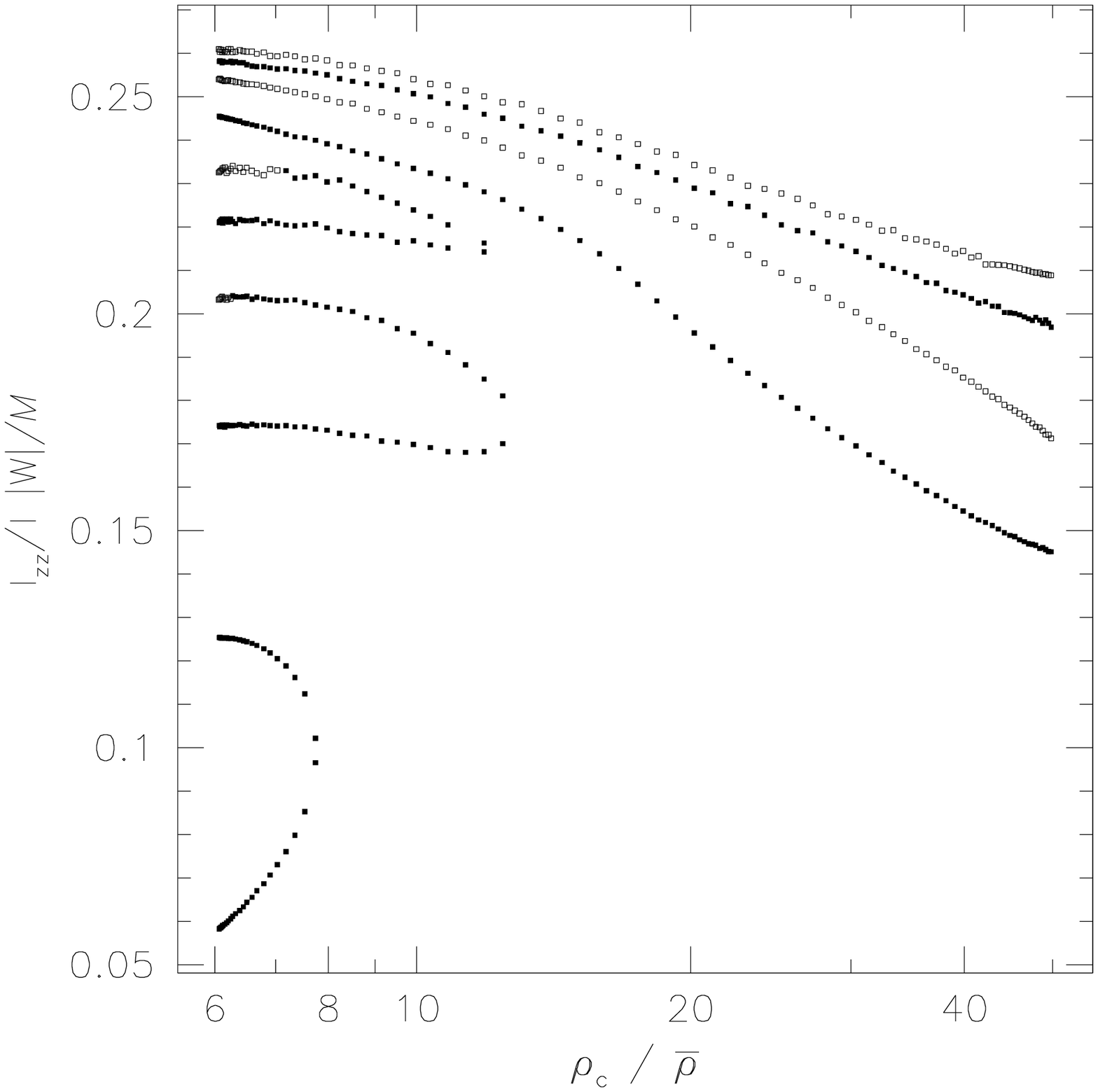}{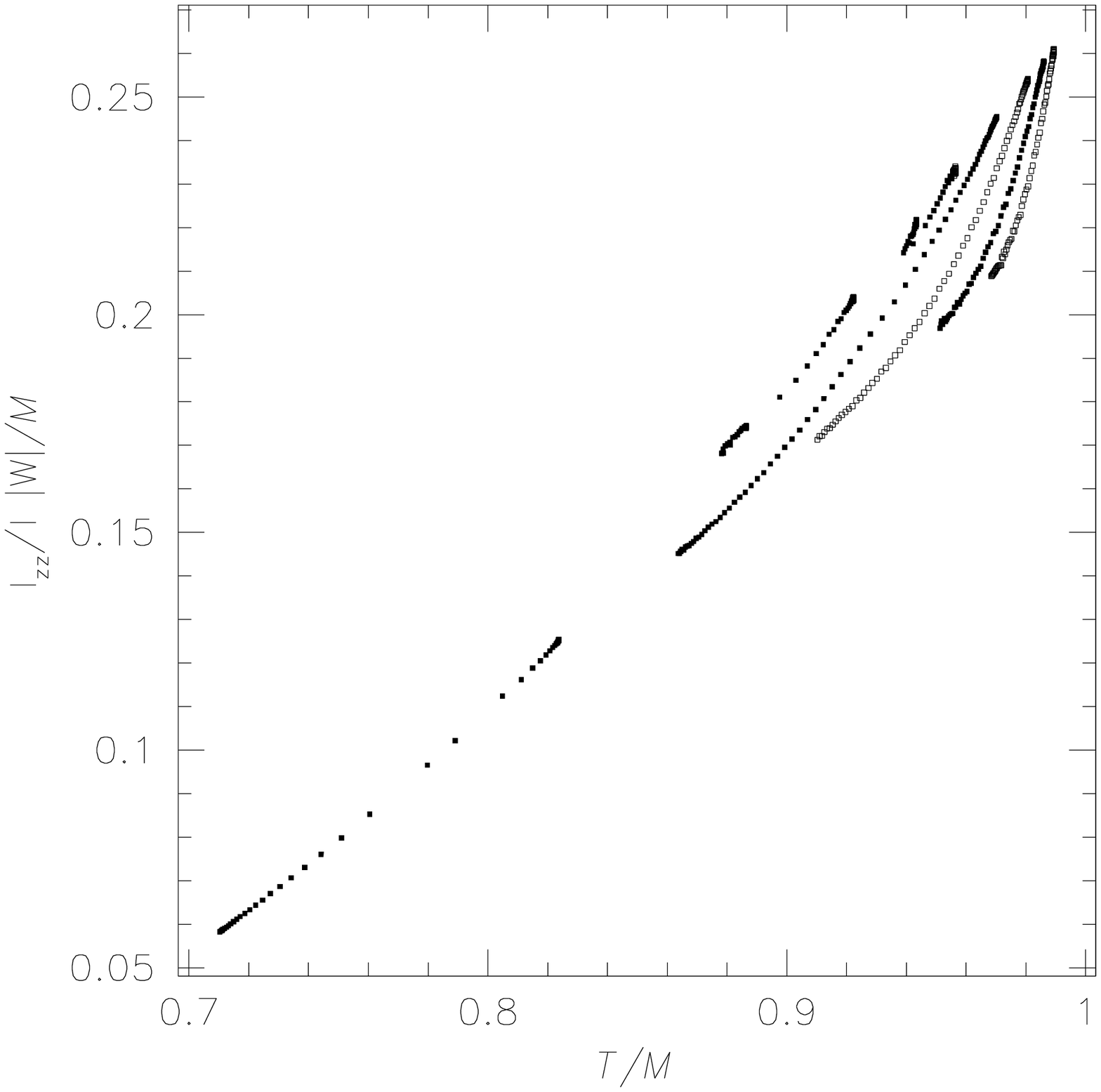}
\caption[]{The reduced quadrupole moment $\ired_{zz}$ as a function
of central density ($\rho_c$) and magnetic field structure. 
${\bar \rho}$ is the mean density of the white dwarf,
$W$ is the total gravitational potential energy of the star and
${\cal M}$ is the total energy of the internal magnetic field.
${\cal T}$ is the toroidal component of the magnetic energy.  Each point 
represents an eigenstate of the coupled magnetic and stellar structure equations.
}
\label{fig:models}
\end{figure}
}

Independent of the configuration of the internal magnetic field, the
reduced quadrupole moment is given approximately by
\be
\ired_{zz} \approx 0.15 \frac{{\cal M}}{|W|} I_0
\ee
for allowed eigenstates of the magnetic configuration.
The moment of inertia of the unperturbed star
($I_0$) decreases from $0.2 M R^2$ to $0.08 M R^2$ as the model
becomes more centrally concentrated.

The value of $\ired_{zz}$ differs in sign 
from the estimates (\eqref{vir}~and~\ref{eq:pert}).  The surface
boundary condition forces the field to have a large toroidal component
that tends to make the star prolate.   The results which do satisfy the
boundary condition may be extrapolated linearly to examine fully
poloidal configurations (which have surface currents).  A fully
poloidal configuration has
\be
\ired_{zz} \approx  -0.5 \frac{{\cal M}}{|W|} I_0 
\ee
while a fully toroidal field has
\be
\ired_{zz} \approx  0.25 \frac{{\cal M}}{|W|} I_0.
\ee
These distortions fall short of the simple estimates by an order of
magnitude because the magnetic field is centrally concentrated, weakening
its effect on the outer layers which dominate the moment of inertia.

If the magnetic field is not aligned with the rotation axis of
the star, the star will emit gravitational radiation.  If the symmetry axis 
of the field makes an angle $\alpha$ with the rotation axis, the components
of the gravitational wave at the Earth are
\ba
h_{+} &=& h_0 \sin \alpha \left \{ 
\frac{1}{2} \cos \alpha \sin i \cos i \cos \left [ \Omega (t - t_0) \right ]
 - \sin \alpha \cos^2 (2i) \cos \left [ 2 \Omega (t-t_0) \right ] 
\right \} \\
h_{\times} &=& h_0 \sin \alpha \left \{ 
\frac{1}{2} \cos \alpha \sin i \cos i \sin \left [ \Omega (t - t_0) \right ]
 - \sin \alpha \cos i \sin \left [ 2 \Omega (t-t_0) \right ] 
\right \} \\
h_0 &=& -\frac{6 G}{c^4} \ired_{zz} \frac{\Omega^2}{d}
\ea
where the inclination between the rotation axis and the line of sight is $i$,
and $d$ is the distance from the star to the detector 
(\cite{1996A&A...312..675B}). 

\section{Potential Sources}

White dwarfs with strong internal fields may have no surface field.
However, it is more likely that strong internal magnetic fields lurk inside
white dwarfs with large external fields.  AM~Her and DQ~Her CVs
consist of a magnetic white dwarf accreting from a late main-sequence
companion.  In AM~Her stars, the accreted material flows directly onto
the magnetic field lines of the degenerate star (\eg
\cite{1990SSRv...54..195C}).  The DQ~ Her stars are the white-dwarf
analogue to low-mass X-ray binaries in which the accreted material
forms a disk around the primary before flowing into the white dwarf
magnetosphere (\eg \cite{1994PASP..106..209P}).  The white dwarfs in
these two types of binaries rotate with periods ranging from tens of
seconds to several hours.  The strength of the surface magnetic field
(for those stars where the magnetic field has been determined) ranges
from 10-100 MG.

\jcite{1998PASP..110.1132P} and \jcite{1999MNRAS.305..473L} argue that
the dwarf nova WZ Sge is indeed a DQ Her star during quiescence.  As
it is thought to be relatively nearby (\cite{1998A&A...339..507M}),
and it is the fastest rotating putative member of the DQ Her class of
magnetic CVs, it may be an important source of gravitational radiation.

About two percent of non-accreting white dwarfs are strongly
magnetized with surface magnetic fields ranging from several tens of
kilogauss to one gigagauss (\cite{Schm95}, \cite{Putn97},
\cite{1999PASP..111..702A}).  The periods of these objects range from
a dozen minutes to over a century, making them a disparate class.
The more strongly magnetized and quickly rotating members of this
group may be important gravitational wave sources.

\figcomment{
\begin{table}
\caption[]{Potential White Dwarf Sources of Gravitational Radiation:
Nearby magnetized white dwarfs with rotational periods less than
3.5$^h$.  The data on the various types of sources was culled from: 
\jcite{1990SSRv...54..195C} for AM Her objects,
\jcite{1994PASP..106..209P} for
DQ Her objects
and \jcite{1999PASP..111..702A} for IWD.  The data for RE~J0317+853 are from
\jcite{1999ApJ...510L..37B} and for WZ Sge are from
\jcite{1998A&A...339..507M}
The distances to AE Aqr and V471 Tau are
from \jcite{1997A&A...323L..49P}.}
\label{tab:srclist}
\medskip
\begin{center}
\begin{tabular}{l|rrrr|l}
\hline
\multicolumn{1}{c|}{Source} &
\multicolumn{1}{c}{Period [s]} &
\multicolumn{1}{c}{Distance [pc]} &
\multicolumn{1}{c}{$B_s$ [MG]} &
\multicolumn{1}{c}{$h$} &
\multicolumn{1}{|c}{Type} \\
\hline
WZ Sge & 28 & 50 & - & $4 \times 10^{-23}$ & Dwarf Nova \\
AE Aqr & 33 & 102 & 50 & $1 \times 10^{-23}$ & DQ Her \\
V533 Her & 64 & 1000 & $-$ & $4 \times 10^{-25}$ & DQ Her \\
DQ Her & 142 & 420 & $-$ & $2 \times 10^{-25}$ & DQ Her \\
H 0253+193 & 206 & 200 & $-$ & $2 \times 10^{-25}$ & DQ Her \\
GK Per & 351 & 490 & $-$ & $3 \times 10^{-26}$ & DQ Her \\
YY Dra & 529 & 155 & $-$ & $4 \times 10^{-26}$ & DQ Her \\
V471 Tau & 555 & 46 & $-$ & $1 \times 10^{-25}$ & DQ Her \\
RE J0317+853 & 726 & 35 & 660 & $1 \times 10^{-25}$ & IWD  \\
V1223 Sgr & 746 & 400 & $-$ & $8 \times 10^{-27}$ & DQ Her \\
AO Psc & 805 & 250 & $-$ & $1 \times 10^{-26}$ & DQ Her \\
RE 0751+144 & 834 & 400 & $-$ & $6 \times 10^{-27}$ & DQ Her \\
BG CMi & 913 & 500 & $-$ & $4 \times 10^{-27}$ & DQ Her \\
FO Aqr & 1254 & 300 & $-$ & $4 \times 10^{-27}$ & DQ Her \\
TV Col & 1910 & 400 & $-$ & $1 \times 10^{-27}$ & DQ Her \\
TX Col & 1911 & 500 & $-$ & $1 \times 10^{-27}$ & DQ Her \\
VZ Pyx & 2918 & 250 & $-$ & $9 \times 10^{-28}$ & DQ Her \\
V1062 Tau & 3726 & 1100 & $-$ & $1 \times 10^{-28}$ & DQ Her \\
EX Hya & 4022 & 100 & $-$ & $1 \times 10^{-27}$ & DQ Her \\
PG 1015+14 & 5940 & 66 & 160 & $8 \times 10^{-28}$ & IWD  \\
VV Pup & 6024 & 145 & 32 & $3 \times 10^{-28}$ & AM Her \\
V834 Cen & 6090 & 86 & 22 & $6 \times 10^{-28}$ & AM Her \\
MR Ser & 6816 & 112 & 25 & $3 \times 10^{-28}$ & AM Her \\
BL Hyi & 6816 & 128 & 30 & $3 \times 10^{-28}$ & AM Her \\
ST LMi & 6834 & 128 & 16 & $3 \times 10^{-28}$ & AM Her \\
EXO 023432-5232.3 & 6876 & 500 & $-$ & $8 \times 10^{-29}$ & AM Her \\
TW Pic & 7188 & 500 & $-$ & $7 \times 10^{-29}$ & DQ Her \\
EXO 033319-2554.2 & 7590 & 250 & 56 & $1 \times 10^{-28}$ & AM Her \\
Feige 7 & 7920 & 49 & 35 & $6 \times 10^{-28}$ & IWD  \\
AM Her & 11136 & 75 & 13 & $2 \times 10^{-28}$ & AM Her \\
V1500 Cyg & 12060 & 1200 & $-$ & $1 \times 10^{-29}$ & AM Her \\
PG 1031+234 & 12240 & 142 & 1000 & $9 \times 10^{-29}$ & IWD  \\

\end{tabular}
\end{center}
\end{table}

\figcomment{
\begin{figure}
\plotone{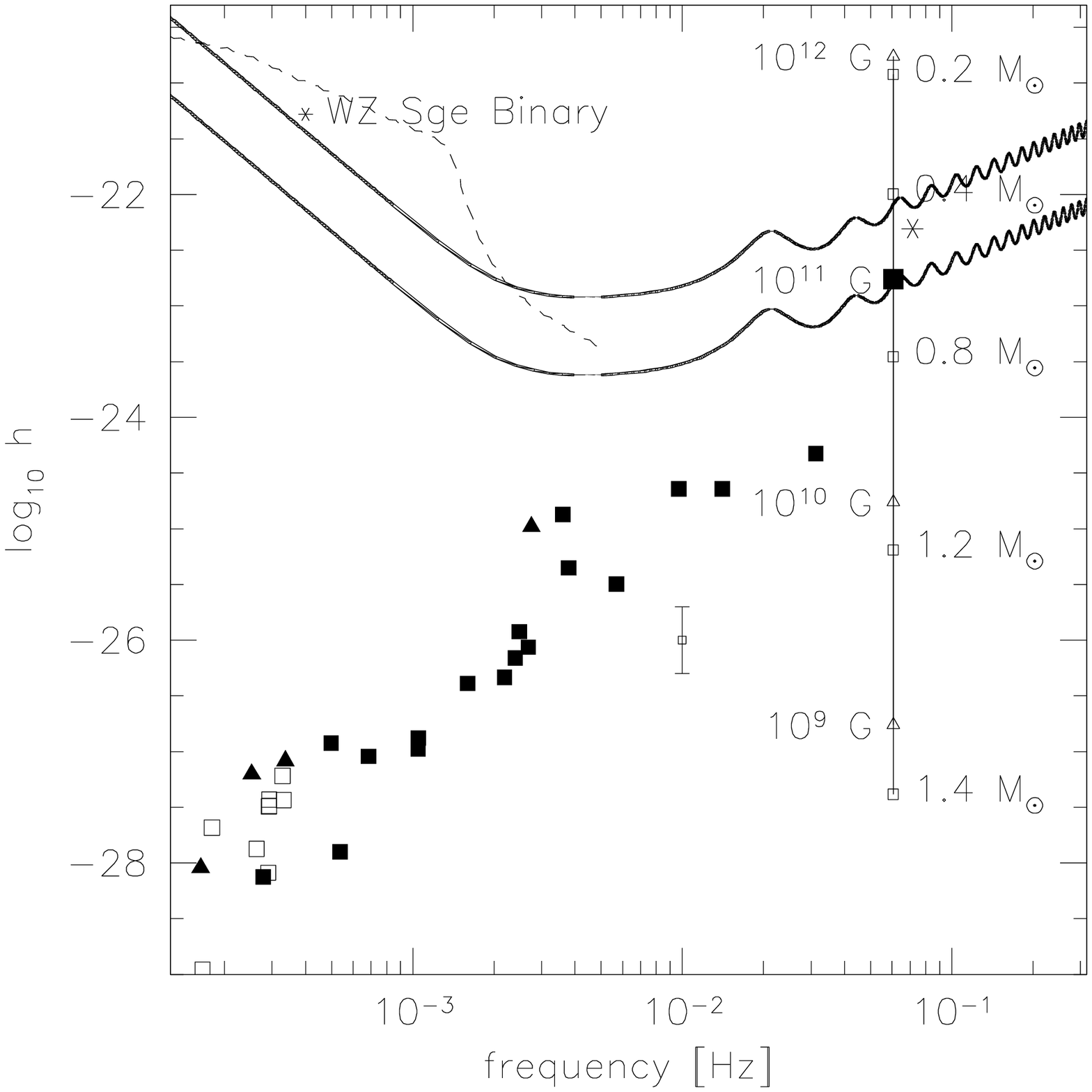}
\caption[]{The estimated strain amplitude ($h$) of gravitational radiation
from nearby white dwarfs.  The solid squares are DQ Her variables, 
the empty squares are AM Her variables, the triangles are isolated 
magnetic white dwarfs and the asterisk is the dwarf nova/DQ Her WZ Sge.  The
second strongest source is AE Aqr which has a legend that
depicts the scaling with the mass of the white dwarf on the right and
with the root-mean-squared magnetic field strength on the left.
The errorbars denote the theoretical uncertainty in $h$ assuming that
mass, radius, total magnetic energy and distance of the white dwarf
are known precisely.  The lower curve shows the sky-averaged noise 
level anticipated for a one-year LISA observation.  The upper curve 
denotes a signal-to-noise ratio of five for a one-year observation.
The dashed curve gives the binary confusion noise estimate (\cite{LISADOC}).}
\label{fig:sources}
\end{figure}
}
}

\tabref{srclist} lists the potential sources along with an estimate of
the strength of the gravitational radiation at the Earth assuming that
the white dwarfs have masses of $0.6 \rmmat{M}_\odot$, the their
RMS internal field is $10^{11}$~G and $\alpha=\pi/2$.  This yields 
values of ${\cal M}/|W| = 4 \times 10^{-3}$ and $\ired_{zz} =
6 \times 10^{-4} I_0 = 1 \times 10^{47}$ g cm$^2$.
The isolated (non-accreting) white dwarfs are denoted by IWD.

\figref{sources} compares the expected gravitational wave amplitudes
for these sources with recent estimates for LISA of instrument thresholds
for achieving a signal-to-noise ratio of unity and five over a
one-year observation (\cite{LISADOC}).  Also, an estimate of the
binary confusion noise and the strength of the radiation from the
binary WZ Sge are depicted.  Unless the internal fields of most of the
potential sources are much larger than $10^{11}$~G, or they are much
less massive that $0.6 \rmmat{M}_\odot$, they will not be detected by
LISA.  However, the two fastest rotating white dwarfs in the sample,
WZ Sge and AE Aqr lie above the signal-to-noise unity line, so these
objects may be detectable after a single year's integration.  In fact,
WZ Sge may appear at two frequencies corresponding to its orbital and
rotational periods.

\section{Discussion}

The internal magnetic fields of white dwarfs are difficult to
constrain observationally or theoretically.  Since the magnetic field
may vanish at the surface, it may leave no obvious trace in the
observed spectra of the star and only affect the global properties of
the star, such as its moment of inertia.  The frequencies of the
pulsation modes of the star may also depend on the structure and
strength of the magnetic field (\cite{1989ApJ...336..403J},
\cite{1995MNRAS.275L..11L}).  However, asteroseismological kernels for
white dwarfs tend to have little amplitude in the central regions of
the star.  If a mode has a node within a composition transition
region, it is pinched between the interface and the surface, and its
amplitude and frequency depends little on the properties of the
material below the interface (\cite{Kawa97}).  Furthermore, unless the
white dwarf is hot, the interior is nearly degenerate, and the
Brunt-V{\"a}is{\"a}l{\"a} frequency almost vanishes, so the observed
g-modes are damped (\cite{Hans94}, \cite{1999ApJ...516..349K}) through
in the interior.  These effects make it difficult to constrain the
magnetic field strength in white dwarf interiors through
asteroseismology.

The strong fields of magnetic white dwarfs may be remnant of the
magnetic fields of their progenitors on the main sequence or produced
subsequently.  As the core of the star forms a white
dwarf, if the magnetic flux is conserved, the resulting remnant will
have a similar ratio of magnetic to gravitational energy
(\cite{1968ApJ...153..797O}); therefore, if the fields are relic, one
can estimate their strength by studying stars along the main sequence.

In main sequence stars it is difficult to exclude the presence of
magnetic fields with a small fraction of the binding energy of the
star.  Helioseismological measurements of the sun place an upper limit
on its internal magnetic field (\cite{1996AAS...188.5307G}).  The
magnetic pressure throughout the solar interior is less than
one-thousandth of the gas pressure, \ie ${\cal M}/|W| < 3 \times
10^{-4}$.  If the magnetic white dwarfs discussed here had solar-like
progenitors, and their fields are relic, the expected signal would be
at least a factor of ten lower than depicted in
\figref{sources}.

However, magnetic white dwarfs may have descended from a
distinct population of main sequence stars, possibly magnetically active
Ap stars (\eg \cite{1981ApJS...45..457A}) which have surface fields of
$10^3 - 10^5$~G and may harbor stronger internal fields than the sun.
\jcite{1985ApJ...296L..27D} argue that the frequency
splittings of the modes rapidly oscillating Ap stars are proportional to the
ratio of the magnetic to the gas pressure averaged over the excited
region of the star.  

Both the magnetic field and the rotation of the star lift the
degeneracy of modes with the same value of $l$ but different values of
$m$.  By comparing the amplitude of each component of the perturbed
modes and assuming a value of the rotational splitting constant
(\cite{1951ApJ...114..373L}), \jcite{1991PASP..103....5M} find $\left
< P_\rmscr{mag}/P_\rmscr{gas} \right > \sim 10^{-3}$.  However,
\jcite{1996ApJ...458..338D} argue that the frequency splitting for the
modes typically measured depends on the strength of the magnetic field
in the outermost two percent radially of the star.  To measure the global
value of $\left < P_\rmscr{mag}/P_\rmscr{gas} \right >$ and thereby
obtain an estimate of ${\cal M}/|W|$, one needs to study an ensemble
of modes which together excite the star globally.  Thus a similar
difficulty arises in estimating the internal fields of the progenitors
of magnetic white dwarfs as did for the white dwarfs themselves.

Whatever has imparted an unusually strong magnetic field on the
surfaces of the magnetic white dwarfs is likely to have boosted their
internal fields as well.  The measurement of gravitational radiation
from rapidly rotating white dwarfs is a straightforward way to
constrain the internal fields of these stars and provide a
clue as to how their strong fields form.

\acknowledgements

The author would like to thank Yu. Levin and E. S. Phinney for useful
discussions and J. M. McDonald and L. Hernquist for providing comments
on the manuscript.  The author received support from a Lee A. DuBridge
postdoctoral fellowship.

\bibliography{wd2,wd,ns,gr,physics}

\begin{thebibliography}{}

\bibitem[\protect{{Andersson}, {Kokkotas} \& {Schutz}~\protect\oyear
  1999\protect\cyear}]{1999ApJ...510..846A}
{Andersson}, N., {Kokkotas}, K. \& {Schutz}, B.~F. 1999,
\newblock {\em \apj,} {\bf 510}, 846.

\bibitem[\protect{{Angel}, {Borra} \& {Landstreet}~\protect\oyear
  1981\protect\cyear}]{1981ApJS...45..457A}
{Angel}, J. R.~P., {Borra}, E.~F. \& {Landstreet}, J.~D. 1981,
\newblock {\em \apjs,} {\bf 45}, 457.

\bibitem[\protect{{Anselowitz} et~al.~\protect\oyear
  1999\protect\cyear}]{1999PASP..111..702A}
{Anselowitz}, T., {Wasatonic}, R., {Matthews}, K., {Sion}, E.~M. \& {MCCook},
  G.~P. 1999,
\newblock {\em \pasp,} {\bf 111}, 702.

\bibitem[\protect{Bender et~al.~\protect\oyear 1998\protect\cyear}]{LISADOC}
Bender, P. {\it et~al.} 1998,
\newblock {\em LISA : Pre-Phase A Report}

\bibitem[\protect{Binney \& Tremaine~\protect\oyear
  1987\protect\cyear}]{Binn87}
Binney, J. \& Tremaine, S. 1987,
\newblock {\em Galactic Dynamics},
\newblock Princeton Univ. Press, Princeton

\bibitem[\protect{{Bonazzola} \& {Gourgoulhon}~\protect\oyear
  1996\protect\cyear}]{1996A&A...312..675B}
{Bonazzola}, S. \& {Gourgoulhon}, E. 1996,
\newblock {\em \aap,} {\bf 312}, 675.

\bibitem[\protect{Brady et~al.~\protect\oyear 1998\protect\cyear}]{Brad98}
Brady, P.~R., Creighton, T., Cutler, C. \& Schutz, B.~F. 1998,
\newblock {\em Phys. Rev. D,} {\bf 57}, 2101.

\bibitem[\protect{{Burleigh}, {Jordan} \& {Schweizer}~\protect\oyear
  1999\protect\cyear}]{1999ApJ...510L..37B}
{Burleigh}, M.~R., {Jordan}, S. \& {Schweizer}, W. 1999,
\newblock {\em \apjl,} {\bf 510}, L37.

\bibitem[\protect{Chandrasekhar~\protect\oyear 1939\protect\cyear}]{Chan39}
Chandrasekhar, S. 1939,
\newblock {\em An Introduction to the Study of Stellar Structure},
\newblock University of Chicago Press, Chicago, Illinois

\bibitem[\protect{{Chandrasekhar} \& {Fermi}~\protect\oyear
  1953\protect\cyear}]{1953ApJ...118..116C}
{Chandrasekhar}, S. \& {Fermi}, E. 1953,
\newblock {\em \apj,} {\bf 118}, 116+.

\bibitem[\protect{{Cropper}~\protect\oyear
  1990\protect\cyear}]{1990SSRv...54..195C}
{Cropper}, M. 1990,
\newblock {\em Space Science Reviews,} {\bf 54}, 195.

\bibitem[\protect{{Cutler} \& {Lindblom}~\protect\oyear
  1987\protect\cyear}]{1987ApJ...314..234C}
{Cutler}, C. \& {Lindblom}, L. 1987,
\newblock {\em \apj,} {\bf 314}, 234.

\bibitem[\protect{{Dziembowski} \& {Goode}~\protect\oyear
  1985\protect\cyear}]{1985ApJ...296L..27D}
{Dziembowski}, W. \& {Goode}, P.~R. 1985,
\newblock {\em \apjl,} {\bf 296}, L27.

\bibitem[\protect{{Dziembowski} \& {Goode}~\protect\oyear
  1996\protect\cyear}]{1996ApJ...458..338D}
{Dziembowski}, W.~A. \& {Goode}, P.~R. 1996,
\newblock {\em \apj,} {\bf 458}, 338+.

\bibitem[\protect{{Friedman}~\protect\oyear
  1983\protect\cyear}]{1983PhRvL..51...11F}
{Friedman}, J.~L. 1983,
\newblock {\em Physical Review Letters,} {\bf 51}, 11.

\bibitem[\protect{{Goode} et~al.~\protect\oyear
  1996\protect\cyear}]{1996AAS...188.5307G}
{Goode}, P.~R., {Dziembowski}, W.~A., {Rhodes}, E.~J., J., {Tomczyk}, S.,
  {Schou}, J. \& {Gong Magnetic Effects Team} 1996,
\newblock {\em American Astronomical Society Meeting,} {\bf 188}, 5307+.

\bibitem[\protect{{Hamada} \& {Salpeter}~\protect\oyear
  1961\protect\cyear}]{1961ApJ...134..683H}
{Hamada}, T. \& {Salpeter}, E.~E. 1961,
\newblock {\em \apj,} {\bf 134}, 683+.

\bibitem[\protect{Hansen \& Kawaler~\protect\oyear 1994\protect\cyear}]{Hans94}
Hansen, C.~J. \& Kawaler, S.~D. 1994,
\newblock {\em Stellar Interiors},
\newblock Springer-Verlag, Berlin

\bibitem[\protect{{Jones} et~al.~\protect\oyear
  1989\protect\cyear}]{1989ApJ...336..403J}
{Jones}, P.~W., {Hansen}, C.~J., {Pesnell}, W.~D. \& {Kawaler}, S.~D. 1989,
\newblock {\em \apj,} {\bf 336}, 403.

\bibitem[\protect{Kawaler, Novikov \& Srinivasan~\protect\oyear
  1997\protect\cyear}]{Kawa97}
Kawaler, S.~D., Novikov, I. \& Srinivasan, G. 1997,
\newblock {\em Stellar Remnants}, Vol.~25 of {\em Saas-Fee advanced course},
\newblock Springer-Verlag, Berlin

\bibitem[\protect{{Kawaler}, {Sekii} \& {Gough}~\protect\oyear
  1999\protect\cyear}]{1999ApJ...516..349K}
{Kawaler}, S.~D., {Sekii}, T. \& {Gough}, D. 1999,
\newblock {\em \apj,} {\bf 516}, 349.

\bibitem[\protect{{Lasota}, {Kuulkers} \& {Charles}~\protect\oyear
  1999\protect\cyear}]{1999MNRAS.305..473L}
{Lasota}, J.~P., {Kuulkers}, E. \& {Charles}, P. 1999,
\newblock {\em \mnras,} {\bf 305}, 473.

\bibitem[\protect{{Ledoux}~\protect\oyear
  1951\protect\cyear}]{1951ApJ...114..373L}
{Ledoux}, P. 1951,
\newblock {\em \apj,} {\bf 114}, 373+.

\bibitem[\protect{{Lindblom}, {Owen} \& {Morsink}~\protect\oyear
  1998\protect\cyear}]{1998PRL..80.4843L}
{Lindblom}, L., {Owen}, B.~J. \& {Morsink}, S.~M. 1998, 
{\em Phys. Rev. Lett.,} {\bf 80}, 4843.

\bibitem[\protect{{Lindblom}~\protect\oyear
  1999\protect\cyear}]{1999gr.qc.....3042L}
{Lindblom}, L. 1999, {\em Phys. Rev. D,} {\bf 60}, 64007.

\bibitem[\protect{{Lou}~\protect\oyear  
  1995\protect\cyear}]{1995MNRAS.275L..11L}
{Lou}, Y. 1995,
\newblock {\em \mnras,} {\bf 275}, L11.

\bibitem[\protect{{Matthews}~\protect\oyear
  1991\protect\cyear}]{1991PASP..103....5M}
{Matthews}, J.~M. 1991,
\newblock {\em \pasp,} {\bf 103}, 5.

\bibitem[\protect{{Meyer-Hofmeister}, {Meyer} \& {Liu}~\protect\oyear
  1998\protect\cyear}]{1998A&A...339..507M}
{Meyer-Hofmeister}, E., {Meyer}, F. \& {Liu}, B.~F. 1998,
\newblock {\em \aap,} {\bf 339}, 507.

\bibitem[\protect{Misner, Thorne \& Wheeler~\protect\oyear
  1973\protect\cyear}]{Misn73}
Misner, C., Thorne, K.~S. \& Wheeler, J.~A. 1973,
\newblock {\em Gravitation},
\newblock W. H. Freeman, San Francisco


\bibitem[\protect{{Ostriker} \& {Hartwick}~\protect\oyear
  1968\protect\cyear}]{1968ApJ...153..797O}
{Ostriker}, J.~P. \& {Hartwick}, F. D.~A. 1968,
\newblock {\em \apj,} {\bf 153}, 797+.

\bibitem[\protect{{Owen} et~al.~\protect\oyear
  1998\protect\cyear}]{1998PRD..58.084020O}
{Owen}, B.~J., {Lindblom}, L., {Cutler}, C., {Schutz}, B.~J., 
{Vecchio}, A. \&
Andersson, N. 1998, 
\newblock {\em Phys. Rev. D,} {\bf 58}, 084020.


\bibitem[\protect{{Patterson}~\protect\oyear
  1994\protect\cyear}]{1994PASP..106..209P}
{Patterson}, J. 1994,
\newblock {\em \pasp,} {\bf 106}, 209.

\bibitem[\protect{{Patterson}~\protect\oyear
  1998\protect\cyear}]{1998PASP..110.1132P}
{Patterson}, J. 1998,
\newblock {\em \pasp,} {\bf 110}, 1132.

\bibitem[\protect{{Perryman} et~al.~\protect\oyear
  1997\protect\cyear}]{1997A&A...323L..49P}
{Perryman}, M. A.~C. {\it et~al.} 1997,
\newblock {\em \aap,} {\bf 323}, L49.

\bibitem[\protect{Putney~\protect\oyear 1996\protect\cyear}]{Putn97}
Putney, A. 1996,
\newblock {\em ApJS,} {\bf 112}, 527.

\bibitem[\protect{Schmidt \& Smith~\protect\oyear 1995\protect\cyear}]{Schm95}
Schmidt, G.~D. \& Smith, P.~S. 1995,
\newblock {\em ApJ,} {\bf 448}, 305.

\bibitem[\protect{Shapiro \& Teukolsky~\protect\oyear
  1983\protect\cyear}]{Shap83}
Shapiro, S.~L. \& Teukolsky, S.~A. 1983,
\newblock {\em Black Holes, White Dwarfs, and Neutron Stars},
\newblock Wiley-Interscience, New York

\bibitem[\protect{{Wentzel}~\protect\oyear
  1961\protect\cyear}]{1961ApJ...133..170W}
{Wentzel}, D.~G. 1961,
\newblock {\em \apj,} {\bf 133}, 170+.

\bibitem[\protect{{Woltjer}~\protect\oyear
  1960\protect\cyear}]{1960ApJ...131..227W}
{Woltjer}, L. 1960,
\newblock {\em \apj,} {\bf 131}, 227+.

\end{thebibliography}
\bibliographystyle{jer}
\figcomment{
\end{document}
\end
}

\singlespace
\begin{table}
\caption[]{Potential White Dwarf Sources of Gravitational Radiation:
Nearby magnetized white dwarfs with rotational periods less than
3.5$^h$.  The data on the various types of sources was culled from: 
\jcite{1990SSRv...54..195C} for AM Her objects,
\jcite{1994PASP..106..209P} for
DQ Her objects
and \jcite{1999PASP..111..702A} for IWD.  The data for RE~J0317+853 are from
\jcite{1999ApJ...510L..37B} and for WZ Sge are from
\jcite{1998A&A...339..507M}
The distances to AE Aqr and V471 Tau are
from \jcite{1997A&A...323L..49P}.}
\label{tab:srclist}
\medskip
\begin{center}
\begin{tabular}{l|rrrr|l}
\hline
\multicolumn{1}{c|}{Source} &
\multicolumn{1}{c}{Period [s]} &
\multicolumn{1}{c}{Distance [pc]} &
\multicolumn{1}{c}{$B_s$ [MG]} &
\multicolumn{1}{c}{$h$} &
\multicolumn{1}{|c}{Type} \\
\hline
WZ Sge & 28 & 50 & - & $4 \times 10^{-23}$ & Dwarf Nova \\
AE Aqr & 33 & 102 & 50 & $1 \times 10^{-23}$ & DQ Her \\
V533 Her & 64 & 1000 & $-$ & $4 \times 10^{-25}$ & DQ Her \\
DQ Her & 142 & 420 & $-$ & $2 \times 10^{-25}$ & DQ Her \\
H 0253+193 & 206 & 200 & $-$ & $2 \times 10^{-25}$ & DQ Her \\
GK Per & 351 & 490 & $-$ & $3 \times 10^{-26}$ & DQ Her \\
YY Dra & 529 & 155 & $-$ & $4 \times 10^{-26}$ & DQ Her \\
V471 Tau & 555 & 46 & $-$ & $1 \times 10^{-25}$ & DQ Her \\
RE J0317+853 & 726 & 35 & 660 & $1 \times 10^{-25}$ & IWD  \\
V1223 Sgr & 746 & 400 & $-$ & $8 \times 10^{-27}$ & DQ Her \\
AO Psc & 805 & 250 & $-$ & $1 \times 10^{-26}$ & DQ Her \\
RE 0751+144 & 834 & 400 & $-$ & $6 \times 10^{-27}$ & DQ Her \\
BG CMi & 913 & 500 & $-$ & $4 \times 10^{-27}$ & DQ Her \\
FO Aqr & 1254 & 300 & $-$ & $4 \times 10^{-27}$ & DQ Her \\
TV Col & 1910 & 400 & $-$ & $1 \times 10^{-27}$ & DQ Her \\
TX Col & 1911 & 500 & $-$ & $1 \times 10^{-27}$ & DQ Her \\
VZ Pyx & 2918 & 250 & $-$ & $9 \times 10^{-28}$ & DQ Her \\
V1062 Tau & 3726 & 1100 & $-$ & $1 \times 10^{-28}$ & DQ Her \\
EX Hya & 4022 & 100 & $-$ & $1 \times 10^{-27}$ & DQ Her \\
PG 1015+14 & 5940 & 66 & 160 & $8 \times 10^{-28}$ & IWD  \\
VV Pup & 6024 & 145 & 32 & $3 \times 10^{-28}$ & AM Her \\
V834 Cen & 6090 & 86 & 22 & $6 \times 10^{-28}$ & AM Her \\
MR Ser & 6816 & 112 & 25 & $3 \times 10^{-28}$ & AM Her \\
BL Hyi & 6816 & 128 & 30 & $3 \times 10^{-28}$ & AM Her \\
ST LMi & 6834 & 128 & 16 & $3 \times 10^{-28}$ & AM Her \\
EXO 023432-5232.3 & 6876 & 500 & $-$ & $8 \times 10^{-29}$ & AM Her \\
TW Pic & 7188 & 500 & $-$ & $7 \times 10^{-29}$ & DQ Her \\
EXO 033319-2554.2 & 7590 & 250 & 56 & $1 \times 10^{-28}$ & AM Her \\
Feige 7 & 7920 & 49 & 35 & $6 \times 10^{-28}$ & IWD  \\
AM Her & 11136 & 75 & 13 & $2 \times 10^{-28}$ & AM Her \\
V1500 Cyg & 12060 & 1200 & $-$ & $1 \times 10^{-29}$ & AM Her \\
PG 1031+234 & 12240 & 142 & 1000 & $9 \times 10^{-29}$ & IWD  \\

\end{tabular}
\end{center}
\end{table}
\clearpage

\doublespace

\begin{figure}
\caption[]{The reduced quadrupole moment $\ired_{zz}$ as a function
of central density ($\rho_c$) and magnetic field structure. 
${\bar \rho}$ is the mean density of the white dwarf,
$W$ is the total gravitational potential energy of the star and
${\cal M}$ is the total energy of the internal magnetic field.
${\cal T}$ is the toroidal component of the magnetic energy.  Each point 
represents an eigenstate of the coupled magnetic and stellar structure equations.
}
\label{fig:models}
\end{figure}

\begin{figure}
\caption[]{The estimated strain amplitude ($h$) of gravitational radiation
from nearby white dwarfs.  The solid squares are DQ Her variables, 
the empty squares are AM Her variables, the triangles are isolated 
magnetic white dwarfs and the asterisk is the dwarf nova/DQ Her WZ Sge.  The
second strongest source is AE Aqr which has a legend that
depicts the scaling with the mass of the white dwarf on the right and
with the root-mean-squared magnetic field strength on the left.
The errorbars denote the theoretical uncertainty in $h$ assuming that
mass, radius, total magnetic energy and distance of the white dwarf
are known precisely.  The lower curve shows the sky-averaged noise 
level anticipated for a one-year LISA observation.  The upper curve 
denotes a signal-to-noise ratio of five for a one-year observation.
The dashed curve gives the binary confusion noise estimate (\cite{LISADOC}).}
\label{fig:sources}
\end{figure}

\clearpage
Figure 1 Left
\medskip
\plotone{rrbiifomo.eps}
\clearpage
Figure 1 Right
\medskip
\plotone{tfiifomo.eps}
\clearpage
Figure 2
\medskip
\plotone{sources.eps}
\end{document}